\documentclass{PoS}

\usepackage{url}

\newcommand{\Fvar}{F_{\rm var}}
\newcommand{\tvar}{t_{\rm var}}
\newcommand{\stat}{_{\rm stat}}

\title{Monitoring of the FSRQ PKS~1510-089 with H.E.S.S.}

\ShortTitle{H.E.S.S. monitoring of PKS~1510-089}

\author{\speaker{Michael Zacharias}$^a$, Felix Jankowsky$^b$, Mahmoud Mohamed$^b$, Heike Prokoph$^c$, David Sanchez$^d$, Stefan Wagner$^b$, Alicja Wierzcholska$^e$ for the H.E.S.S. Collaboration\\ %, Martin Bell et al. for the MWA collaboration
        $^a$Centre for Space Research, North-West University, 2520 Potchefstroom, South Africa\\
        $^b$Landessternwarte, Universit\"at Heidelberg, K\"onigstuhl, 69117 Heidelberg, Germany\\
        $^c$GRAPPA, Anton Pannekoek Institute for Astronomy and Institute of High-Energy Physics, University of Amsterdam, Science Park 904, 1098 XH Amsterdam, The Netherlands\\
        $^d$Laboratoire d'Annecy-le-Vieux de Physique des Particules, Universite Savoie Mont-Blanc, CNRS/IN2P3, 74941 Annecy-le-Vieux, France\\
        $^e$Instytut Fizyki Jadrowej PAN, ul. Radzikowskiego 152, 31-342 Krakow, Poland\\
        E-mail: \email{mzacharias.phys@gmail.com}}

\abstract{The flat spectrum radio quasar (FSRQ) PKS 1510-089 (z=0.361) is known for its complex multiwavelength behavior. It has been monitored regularly at very high energy (VHE, $E>100\,$GeV) gamma-rays with H.E.S.S. since its discovery in 2009 in order to study the unknown behavior of FSRQs in quiescence at VHE, as well as the flux evolution around flaring events. Given the expected strong cooling of electrons and the absorption of VHE emission within the broad-line region, a detection of PKS 1510-089 at VHE in a quiescent state would be an important result, implying an acceleration and emission region on scales beyond the broad-line region. The H.E.S.S. monitoring has been intensified since 2015 and is complemented by monitoring at high energy ($E>100\,$MeV) gamma-rays with Fermi, at X-rays with Swift-XRT, and at optical frequencies with ATOM. The dense lightcurves allow for the first time detailed comparison studies between these energy bands. The source has been active in several frequency bands for a large fraction of the observation time frames. Yet, we do not find obvious correlations between the VHE and the other bands over the observed time frame indicating a non-trivial interplay of the acceleration, cooling and radiative processes. It also implies a rich variety in flaring behavior, which makes this source difficult to interpret within a unique theoretical framework.} %, and in the radio with MWA

\FullConference{35th International Cosmic Ray Conference -- ICRC2017\\
		10-20 July, 2017\\
		Bexco, Busan, Korea}

\begin{document}
\section{Introduction} \label{sec:intro}
Flat spectrum radio quasars (FSRQs) belong to the blazar class of active galactic nuclei (AGN). Blazars host relativistic jets which are closely aligned with the line of sight \cite{br74}, resulting in strongly beamed radiation in the observer's frame. In turn, the detected emission is dominated by the jet over the entire electromagnetic spectrum making blazars prime objects to study the inner workings of relativistic jets.

Blazar spectra are dominated by two broad components, with the low energy one being attributed to electron-synchrotron emission. There is debate about the origin of the high-energy component, and whether it is dominated by non-thermal electron or proton emission. In leptonic models, which assume protons to be a cold background, the emission is of inverse-Compton origin, where the internal synchrotron emission or external soft photon fields can serve as target photons. In hadronic models, proton synchrotron emission dominates the high-energy component with additions of synchrotron emission from secondary particles, like pions, muons, and electron-positron pairs, which might form a cascade.

Naturally, different variability patterns and correlations are expected for different models. Especially, in the simple one-zone leptonic model, strong correlations are expected between all energy bands, since the same particle population is responsible for the entire electromagnetic output. While a change in the external fields might still be able to induce stronger variations in the inverse Compton component than in the synchrotron component, there should still be correlated variability. In the hadronic model, there is not necessarily a strong correlation or any correlation at all expected between the low-energy and the high-energy spectral components, since they are produced from independent particle populations. Hence, the analysis of correlations between different bands might put strong constraints on the simplest models.

This is one of the reasons why long-term monitoring of blazars in various energy bands is essential to understand the inner workings of relativistic jets. H.E.S.S. has conducted such observations on the high-frequency peaked BL Lac object PKS~2155-304, which is an easy-to-detect target in the very high energy (VHE, $E>100\,$GeV) $\gamma$-ray band. Numerous results of this effort have been published \cite{Hea10,Hea14,Hea17}, which reveal that different correlation patterns between the optical and the $\gamma$-ray regime hold at different epochs.

A similar monitoring campaign was started with H.E.S.S. on the FSRQ PKS~1510-089. While it has been observed with H.E.S.S. regularly since its detection in 2009 \cite{Hea13}, the dedicated monitoring effort has been launched in 2015 with several hours of observation time during each observation period. Unlike high-frequency peaked BL Lac objects, FSRQs are difficult to detect with ground-based $\gamma$-ray observatories, since the peak of the $\gamma$-ray component is typically located at around $100\,$MeV. In addition to the potential absorption of VHE $\gamma$-rays by the strong optical nuclear photon fields, this poses difficulties for the detection of the source at VHE.

Here the data and the analysis of the seasons 2009-2015 are presented with a gap of data in 2013 and 2014, where no observations were taken. Observations from Fermi-LAT in the high energy (HE, $100\,$MeV$<E<100\,$GeV) $\gamma$-ray domain were also analyzed. For 2015 additional data has been gathered from Swift-XRT in the X-ray band, and from ATOM in the optical R-band. %radio data from the Murchison Widefiled Array (MWA) are shown, along with

\section{Data analysis} \label{sec:ana}
\subsection{H.E.S.S.}
H.E.S.S. is an array of five Imaging Atmospheric Cherenkov Telescopes located in the Khomas Highland in Namibia at an altitude of about $1800\,$m. The first phase of H.E.S.S. began in 2004 with four 12-m telescopes (CT1-4) giving an optimal energy threshold of $\sim 100\,$GeV. In 2012, a 28-m telescope (CT5) was added to the array, reducing the energy threshold to $\sim 50\,$GeV. The observations performed between 2009 and 2012 were done with CT1-4, while most observations in 2015 were conducted with CT5 in monoscopic mode. Therefore, the 2015 data set has to be analyzed separately from the 2009-2012 data set.

For the observations between 2009 and 2012, a total number of 85 runs (1 run lasts for $28\,$min) passed the standard quality selection \cite{aHea06} resulting in a total live time of $36\,$hrs. While the data taken in 2009 has already been published \cite{Hea13}, it was reanalyzed along with the subsequent years. The data set has been analyzed with the Model analysis chain using \textsc{loose cuts} \cite{dnr09}, and cross-checked using the independent reconstruction and analysis chain ImPACT \cite{ph14}. The source is detected with $9.1\sigma$ with an energy threshold of $146\,$GeV. The observed photon spectrum is well described by a power-law 
\begin{eqnarray}
 F(E) = N(E_0)\times(E/E_0)^{-\Gamma} \label{eq:powlaw},
\end{eqnarray}
with a normalization $N=(10\pm 1_{\rm stat} \pm 2_{\rm sys})\times 10^{-12}\,$cm$^{-2}$s$^{-1}$TeV$^{-1}$, a decorrelation energy $E_0 = 0.265\,$TeV, and a photon index $\Gamma = 3.6\pm 0.3_{\rm stat}$. The systematic errors on the flux normalization are $20\%$. The average photon flux above $150\,$GeV is $(4.7\pm0.6_{\rm stat}\pm0.9_{\rm sys})\times 10^{-12}\,$cm$^{-2}$s$^{-1}$.

For the 2015 season, a total number of 155 runs passed the quality selection, giving a total live time of $66.8\,$hrs. The data set has been analyzed with the Model analysis chain using \textsc{mono standard cuts} \cite{hea15}, and the cross-check was done again with ImPACT \cite{pmg15}. The bright star Beta Libra, located roughly $1\,$deg away from PKS~1510-089, causes an artifact in the mono background maps, which has an influence on the on-source region on the order of $5\%$. The systematic error on the flux normalization has therefore been raised to $25\%$. The energy threshold of the data set is $\sim 80\,$GeV.

In 2015, the source is detected with $15.3\sigma$. The observed photon spectrum is well described by a power-law, with a normalization $N=(250\pm 12_{\rm stat} \pm 62_{\rm sys})\times 10^{-12}\,$cm$^{-2}$s$^{-1}$TeV$^{-1}$, a decorrelation energy $E_0 = 0.132\,$TeV, and a photon index $\Gamma = 3.8\pm 0.1_{\rm stat}$. The average photon flux above $150\,$GeV is $(7.9\pm0.4_{\rm stat}\pm1.9_{\rm sys})\times 10^{-12}\,$cm$^{-2}$s$^{-1}$.

%The nightly averaged lightcurve for all seasons, 2009-2015, is shown in the top panel of Figure \ref{fig:lc_hf}. The average flux above $150\,$GeV is $(5.5\pm0.4_{\rm stat})\times 10^{-12}\,$cm$^{-2}$s$^{-1}$, neglecting the different systematic error levels. A constant flux is ruled out with more than $10\sigma$.

%
\subsection{Fermi-LAT}
Fermi-LAT monitors the HE $\gamma$-ray sky every $3\,$hrs in the energy range from $20\,$MeV to beyond $500\,$GeV \cite{aFea09}. An analysis of the (publicly available) Pass 8 SOURCE class events above an energy of $100\,$MeV was performed for a Region of Interest (RoI) of $15^{\circ}$ radius centered at the position of PKS~1510-089 in the time frame 2009 to 2015. All sources within the RoI (and $7^{\circ}$ beyond) pertaining to the 3FGL catalog have been accounted for in the likelihood analysis. In order to reduce contamination from the Earth Limb, a zenith angle cut of $<90^{\circ}$ was applied. The analysis was performed with the ScienceTools software package version v10r0p5 using the \textsc{P8R2\_SOURCE\_V6}\footnote{\url{http://fermi.gsfc.nasa.gov/ssc/data/analysis/documentation/Cicerone/Cicerone_LAT_IRFs/IRF_overview.html}} instrument response function and the \textsc{GLL\_IEM\_v06} and \textsc{ISO\_P8R2\_SOURCE\_V6\_v06} models\footnote{\url{http://fermi.gsfc.nasa.gov/ssc/data/access/lat/BackgroundModels.html}} for the Galactic and isotropic diffuse emission \cite{aFea16a}, respectively.
In order to produce a finely binned ($24\,$hr bins) lightcurve, a power-law model for the spectrum was assumed with the parameters left free to vary from bin to bin.

\subsection{Swift-XRT}
XRT \cite{bea05} on board the Swift satellite is an X-ray detector sensitive in the energy range $0.3-10\,$keV. Observations were carried out in photon counting mode. X-ray data collected in 2015 were analyzed using the HEAsoft software package version 6.16\footnote{\url{http://heasarc.gsfc.nasa.gov/docs/software/lheasoft}} with the latest CALDB. The events were cleaned and calibrated using the \verb|xrtpipeline| task and the data in the $0.3$--$10\,$keV energy range with grades 0-12 were analyzed.
The light curve flux points were calculated from the spectra of a single observation integrating between $2$ and $10\,$keV. For the spectral fitting XSPEC v. 12.8.2 with a single power-law model and Galactic hydrogen absorption fixed to the Galactic value of $n_H = 6.99 \cdot10^{20}\,$cm$^{-2}$ \cite{kea05} were used. %More complicated spectra, like a broken power-law or a log-parabola, did not improve the fit. %\cite{a96} 

\subsection{ATOM}
ATOM is a $75\,$cm optical telescope located on the H.E.S.S. site \cite{hea04}. Operating since 2005, ATOM provides optical monitoring of $\gamma$-ray sources and has observed PKS~1510-089 with high cadence in the R band. The magnitudes of each flux point have been derived with differential photometry using five comparison stars in the same field of view. The resulting fluxes of 2015 have been corrected for Galactic extinction.

%
%\subsection{MWA}
%

%
\section{Lightcurves and correlations} \label{sec:corr}
\begin{figure*}[t]
\begin{minipage}{0.49\linewidth}
\vspace{-0.95cm} \centering %\resizebox{\hsize}{!}
{\includegraphics[width=0.8\textwidth]{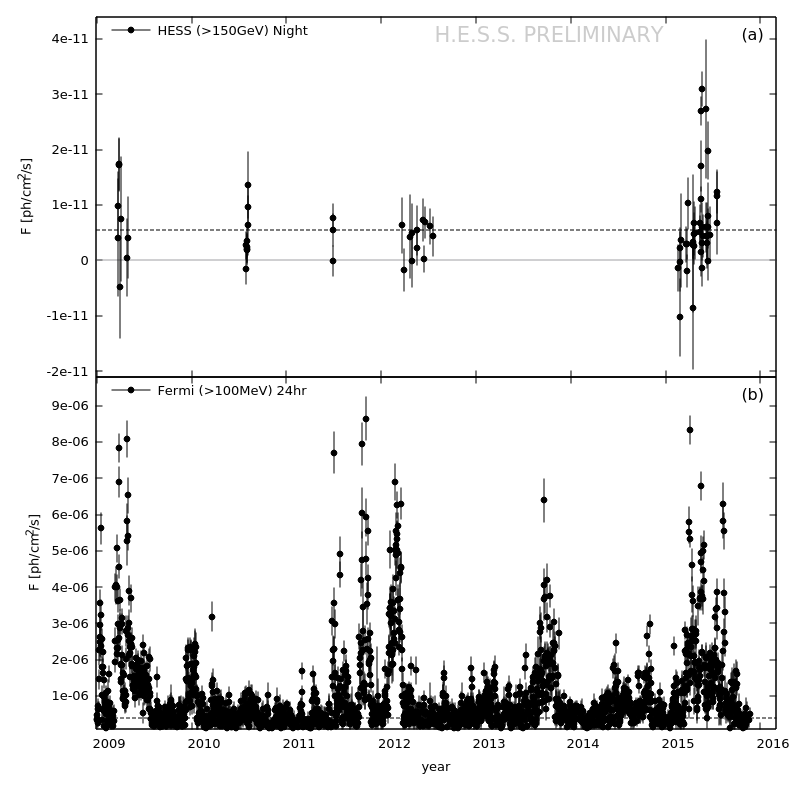}}
\caption{Nightly-averaged lightcurve of {\bf (a)} H.E.S.S. and {\bf (b)} $24\,$hr-averaged lightcurve of Fermi-LAT for all observations between 2009 and 2015. Solid lines mark the Null-level, while dashed lines mark long-term averages.}
\label{fig:lc_hf}
\end{minipage}
\hspace{\fill}
\begin{minipage}{0.49\linewidth}
\centering %\resizebox{\hsize}{!}
{\includegraphics[width=0.8\textwidth]{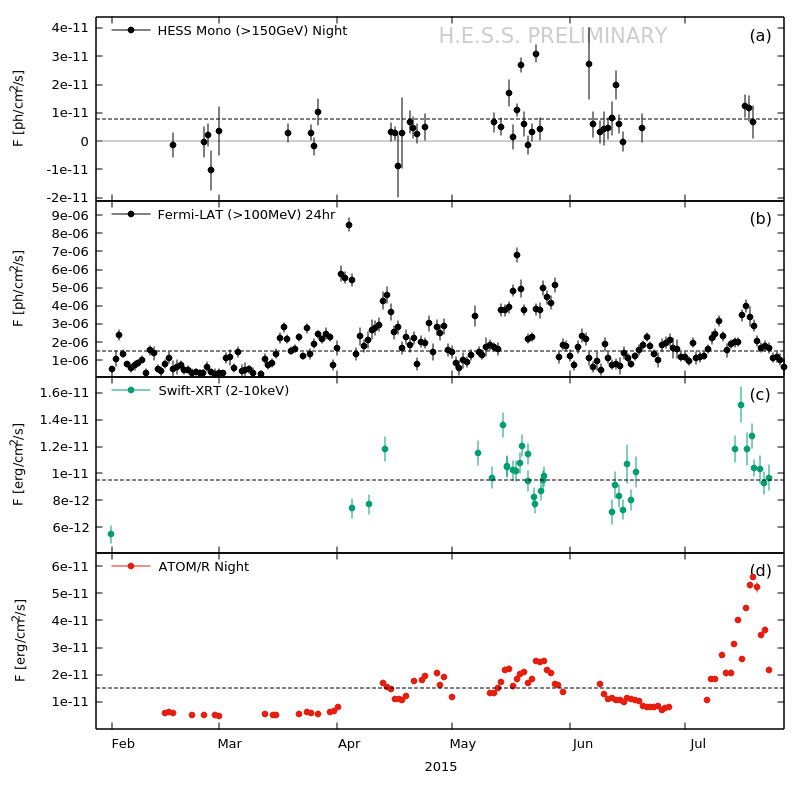}}
\caption{Lightcurves of PKS~1510-089 in 2015. {\bf (a)} Nightly-averaged H.E.S.S. VHE lightcurve. 
{\bf (b)} Fermi-LAT HE lightcurve averaged over $24\,$hrs. 
%{\it Third panel:} Evolution of the HE spectral index assuming a power-law spectrum in $24\,$hr bins. 
{\bf (c)} Swift-XRT X-ray lightcurve. 
{\bf (d)} Nightly-averaged ATOM optical lightcurve. 
%{\it Bottom panel:} Radio lightcurve from MWA. 
In all panels solid lines mark the Null-level, while dashed lines mark the 2015 average.}
\label{fig:mwl_lc_all}
\end{minipage}
\end{figure*} 
%
%
% \begin{figure}[t]
% \centering
% \includegraphics[width=0.48\textwidth]{}
% \caption{Lightcurves of PKS~1510-089 of the observation period. {\it Top panel:} H.E.S.S. VHE lightcurve in nightly averaged. {\it Second panel:} HE lightcurve averaged over $24\,$hrs from Fermi-LAT. {\it Third panel:} Evolution of the HE spectral index assuming a power-law spectrum in $24\,$hr bins. {\it Fourth panel:} X-ray lightcurve from Swift-XRT. {\it Fifth panel:} Optical lightcurve averaged over one night from ATOM. %{\it Bottom panel:} Radio lightcurve from MWA. 
% In all panels solid lines mark the Null-level, while dashed lines mark long-term averages.}
% \label{fig:mwl_lc_all}
% \end{figure}
%
%
\begin{figure*}[t]
\begin{minipage}{0.49\linewidth}
\centering %\resizebox{\hsize}{!}
{\includegraphics[width=0.8\textwidth]{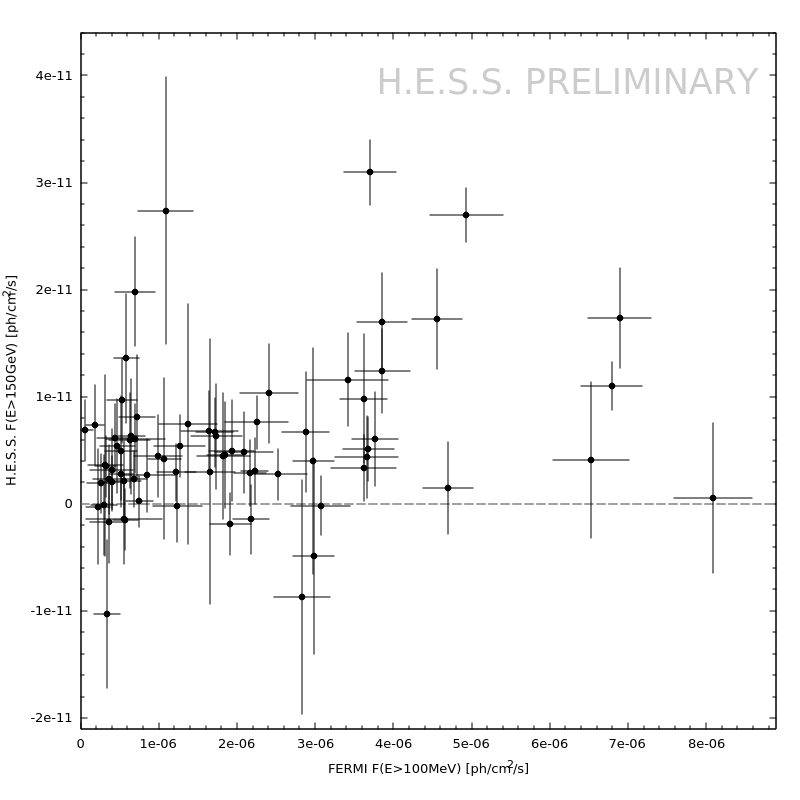}}
\end{minipage}
\hspace{\fill}
\begin{minipage}{0.49\linewidth}
\centering %\resizebox{\hsize}{!}
{\includegraphics[width=0.8\textwidth]{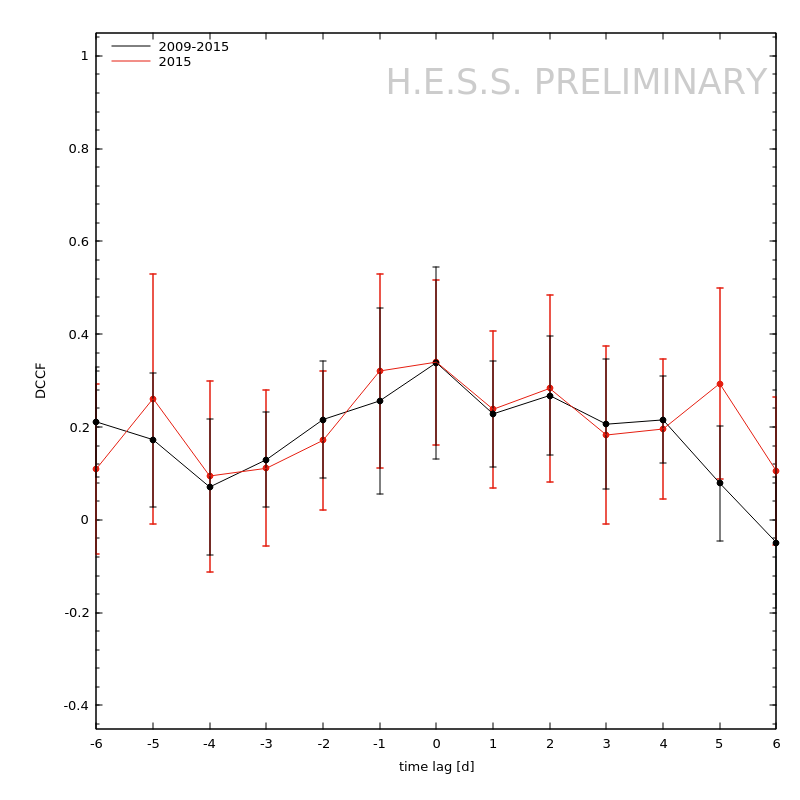}}
\end{minipage}
\caption{Flux-flux scatterplot (left) and DCCF (right) for H.E.S.S. versus Fermi-LAT fluxes for all data between 2009 and 2015. The black symbols in the right plot mark the DCCF using all data, while the red symbols mark the DCCF for the 2015 subset.}
\label{fig:corr_hf}
\end{figure*} 
\begin{figure*}[t]
\begin{minipage}{0.49\linewidth}
\centering %\resizebox{\hsize}{!}
{\includegraphics[width=0.8\textwidth]{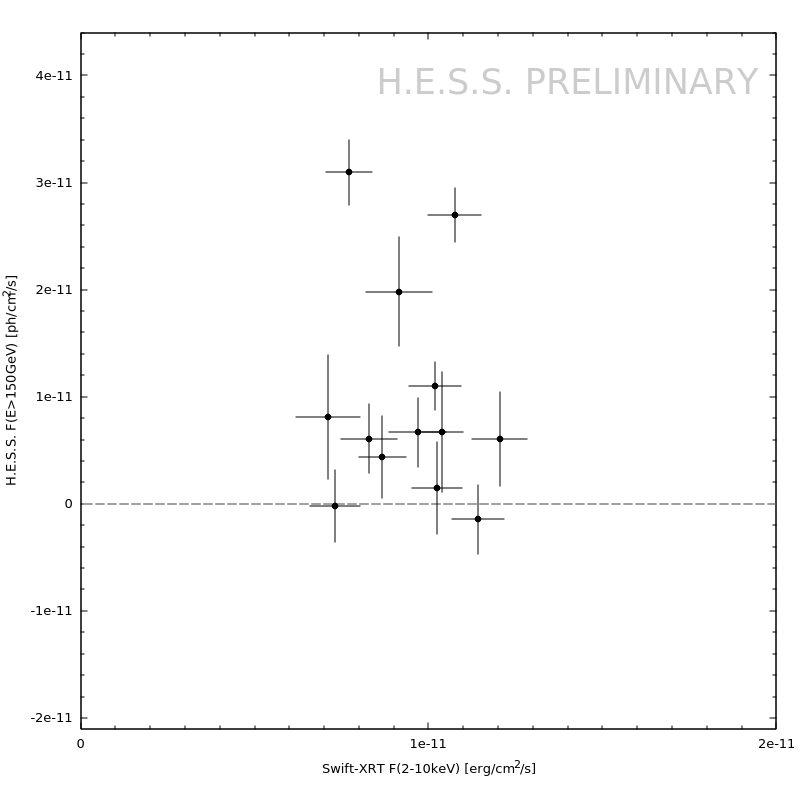}}
\end{minipage}
\hspace{\fill}
\begin{minipage}{0.49\linewidth}
\centering %\resizebox{\hsize}{!}
{\includegraphics[width=0.8\textwidth]{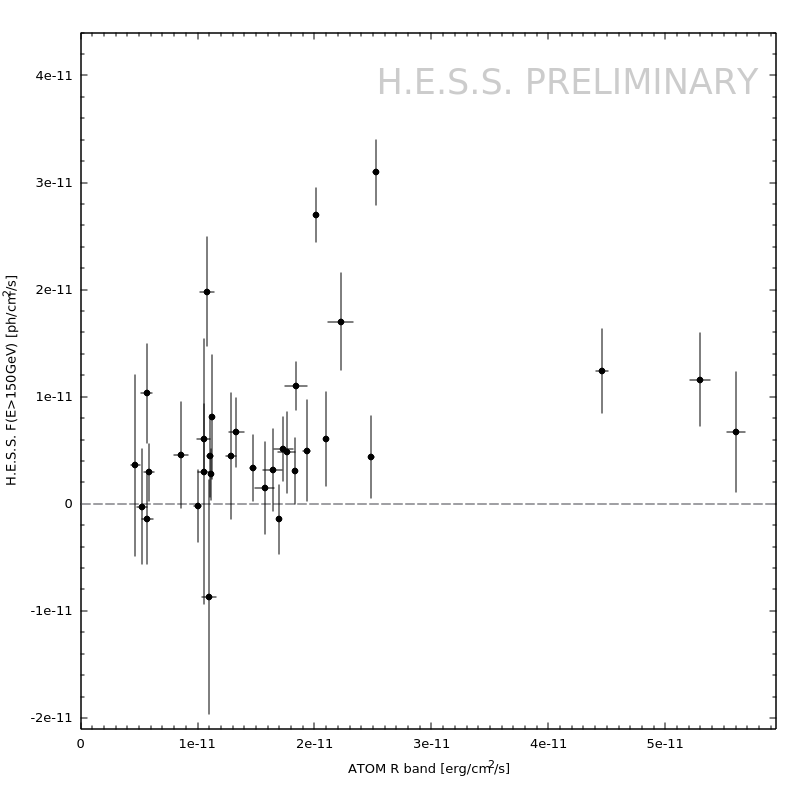}}
\end{minipage}
\caption{Flux-flux scatterplots for H.E.S.S. versus Swift-XRT (left) and ATOM (right) for observations in 2015.}
\label{fig:corr}
\end{figure*} 
The $\gamma$-ray lightcurves observed with H.E.S.S. and Fermi-LAT are shown in Fig. \ref{fig:lc_hf}. The H.E.S.S. lightcurve shows nightly averaged photon fluxes integrated above $150\,$GeV with an average flux of $(5.5\pm0.4_{\rm stat})\times 10^{-12}\,$cm$^{-2}$s$^{-1}$, neglecting the different systematic error levels of the two data sets. A constant flux is ruled out with more than $10\sigma$. The Fermi-LAT fluxes in the bottom panel are averaged over $24\,$hrs and integrated above $100\,$MeV. The HE average photon flux is $(4.01\pm0.03_{\rm stat})\times 10^{-7}\,$cm$^{-2}$s$^{-1}$, but the lightcurve is also highly variable. 
The variability can be further quantified with the fractional variability
\begin{eqnarray}
 \Fvar = \frac{\sqrt{S^2 - \sigma_{\rm err}^2}}{\bar{F}} \label{eq:fvar},
\end{eqnarray}
with the average flux $\bar{F}$ of the considered data set, the mean square error $\sigma_{\rm err}^2$, and the variance $S^2$ \cite{eea02}. The VHE lightcurve is strongly variable with $\Fvar = 1.0\pm 0.2\stat$, as is the HE lightcurve with $\Fvar = 1.164\pm0.005\stat$. If one excludes the 2015 data set in the VHE lightcurve, the fractional variability is reduced to $\Fvar = 0.8\pm 0.4\stat$. While this is still compatible with the $\Fvar$ of the entire data set, the larger error indicates that this value is less significant. This implies that most of the variability comes from the (more detailed) 2015 data set, and that VHE flares could be rare in PKS~1510-089. 

The minimum variability time scale between subsequent measurements is defined as \cite{ww95}
\begin{eqnarray}
 \tvar = \bar{F} \frac{t_{i+1} - t_{i}}{|F_{i+1}-F_{i}|} \label{eq:tvar} .
\end{eqnarray}
The fastest variability in the VHE band occurred in May 2015 with $\tvar = 6\pm1\stat\,$hrs. In the HE band the fastest variability took place in October 2011 with $\tvar = 0.82\pm0.07\stat\,$hrs.

The multiwavelength lightcurves of the 2015 season are shown in Fig. \ref{fig:mwl_lc_all}. Panel (a) shows again the nightly averages of the H.E.S.S. observations. The VHE lightcurve exhibits an $\Fvar = 1.1\pm0.1\stat$, and the same minimum variability time scale as above. %This indicates that most of the variability in the VHE band was indeed observed during 2015.

Fig. \ref{fig:mwl_lc_all}(b) shows the HE lightcurve measured with Fermi-LAT. The HE lightcurve is also very variable with $\Fvar = 0.72\pm 0.01\stat$, even though this value indicates lower variability than for the entire data set. One should note, however, that the average flux is higher in the 2015 season than for the entire data set, which could affect the degree of variability. The fastest variability took place during the flare in early April with $\tvar = 9\pm1\stat\,$hrs, which unfortunately could not be covered by H.E.S.S. due to Moon constraints. %The photon index of the HE band, as displayed in the third panel of Fig. \ref{fig:mwl_lc_all}, is also showing some variability, which might indicate a hardening of the spectrum with flux. Especially, during bright VHE states there seems to be a trend in the HE band to exhibit harder spectra. However, due to a lack of statistics, i.e. a low number of high flux states in the VHE, this cannot be quantified.

Fig. \ref{fig:mwl_lc_all}(c) shows the X-ray flux integrated between $2$ and $10\,$keV as measured with Swift-XRT. Over the shown time period, the flux varies by less than a factor $3$, which can be quantified by a low level in $\Fvar = 0.17\pm0.02\stat$ and $\tvar = 9\pm5\stat\,$hrs, with the latter occurring during the VHE flare in May 2015. While the latter is on the same order of magnitude as the VHE variability, the larger error clearly shows that this time scale is less significant. Given that the X-rays mark the low-energy part of the inverse-Compton component, a lower variability than in the $\gamma$ rays is expected.

The optical R-band, on the other hand, is in the high-energy part of the synchrotron component. The nightly averaged lightcurve, shown in Fig. \ref{fig:mwl_lc_all}(d), is clearly variable with $\Fvar = 0.666\pm0.004\stat$. Despite this strong degree of variability, the variability itself is not as fast as in the $\gamma$-ray component, with a minimum variability time scale constrained to $\tvar = 18.6\pm0.8\stat\,$h. This occurred during the bright optical flare in July 2015.

Interestingly, comparing the lightcurves of the VHE band with the lightcurves of the other bands does not reveal obvious correlations. This can be further quantified by flux-flux scatterplots, which should reveal correlations with no time lag, and the discrete cross-correlation function (DCCF), which can search for correlations with a time lag \cite{ek88}. Since the latter requires a huge data set to be sensitive enough, it was only used for the VHE and HE data set of 2009-2015. The scatterplot in Fig. \ref{fig:corr_hf} does not show any obvious trend. A linear fit to the data, and hence a zero time lag, is ruled out with a p-value less than $10^{-20}$. The DCCF in Fig. \ref{fig:corr_hf} also does not find a significant correlation between the two data sets on the order of a few days, which holds for both the entire data set (shown in black) and the 2015 subset (red).

Since the data sets in 2015 in the VHE, X-ray and R-band are too sparse to compute meaningful DCCFs, only the scatterplots are displayed in Fig. \ref{fig:corr}. The VHE versus X-ray data is too sparse to draw firm conclusions even for zero-lag correlations. On the other hand, the optical flux seems to be enhanced if the VHE flux is enhanced, which however does not work the other way around. Hence, one cannot find a clear correlation between these two bands, either. Not surprisingly, for both scatterplots a linear trend is ruled out with a p-value less than $10^{-20}$.

%In order to search for correlations between the VHE band and the other bands, flux-flux scatterplots have been created, which are displayed in the top row of Fig. \ref{fig:corr}. In neither of the plots, a trend between the VHE flux and the HE, X-ray, or optical flux is obvious. In all cases the fit of a linear regression to the data yields a p-value of less than $10^{-20}$, clearly proving that no (linear) trends are present.

%While flux-flux scatterplots can show obvious trends and correlations between different bands, they might not be a sensitive tool, if time-lags are present. The discrete cross-correlation function (DCCF) by \cite{ek88} has been employed to search for time-lags between the VHE band and the other bands. According to the bottom row of Fig. \ref{fig:corr}, no significant correlations or time-lags are found, strengthening the interesting case that fluxes in the VHE band are not obviously correlated with flux states in other bands.

%
\section{Discussion \& Summary} \label{sec:sum}
PKS~1510-089 is a puzzling source, and correlations between different bands are known to change from flare to flare \cite{bea14,b13,sea13,sea15}. 

VHE observations presented here, were gathered with H.E.S.S. between 2009 and 2012, and in 2015, adding another piece to the puzzle. Contemporaneous multiwavelength data in the HE band from Fermi-LAT (for the entire period), in the X-ray band from Swift-XRT, and in the optical R-band from ATOM (both for 2015) have also been shown. This is the first time that such a dedicated program on an FSRQ is reported. In most covered bands, the source was highly variable with $\Fvar>0.6$, and only the X-rays were less variable with $\Fvar = 0.17\pm0.02\stat$. The minimum variability time scale in the HE and VHE band can be constrained to less than $10\,$hrs, while the X-ray and optical minimum variability time scale is less than a day.

A search for correlation patterns between the VHE band and the other bands was conducted. Interestingly, no correlations could be established, neither through inspections of flux-flux scatterplots, nor through the DCCF. Apparently, VHE flares are rare, and high states in lower energy bands are not necessarily reflected by a high state in the VHE band. Likewise, high states in the VHE band also do not necessarily exhibit a high state in the other bands, as is obvious from the scatterplots in Figs. \ref{fig:corr_hf} and \ref{fig:corr}. There might be an indication that high VHE states are reflected by high states in the optical band, but this does not hold the other way around. The results presented in \cite{zea17} about a strong VHE flare in 2016 also show this behavior.

In turn, the parameters of the processes underlying the flaring events must change from flare to flare, or else correlations between the VHE band and other bands should be obvious. The change in parameters could mean several things. For instance, different locations of the emission region (i.e. within or outside the BLR), changing efficiencies of the acceleration process (i.e. the maximum Lorentz factor particles can reach), or a mixture of both, to name only two obvious choices. It will be an interesting task to find a suitable explanation for the changing behavior in PKS~1510-089.

Naturally, these conclusions are hampered by the fact that the size of the data set is still limited. Fortunately, the monitoring program with H.E.S.S. continues and promises a very rich data set.

\section*{Acknowledgements}
\footnotesize{The support of the Namibian authorities and of the University of Namibia in facilitating the construction and operation of H.E.S.S. is gratefully acknowledged, as is the support by the German Ministry for Education and Research (BMBF), the Max Planck Society, the German Research Foundation (DFG), the Alexander von Humboldt Foundation, the Deutsche Forschungsgemeinschaft, the French Ministry for Research, the CNRS-IN2P3 and the Astroparticle Interdisciplinary Programme of the CNRS, the U.K. Science and Technology Facilities Council (STFC), the IPNP of the Charles University, the Czech Science Foundation, the Polish National Science Centre, the South African Department of Science and Technology and National Research Foundation, the University of Namibia, the National Commission on Research, Science \& Technology of Namibia (NCRST), the Innsbruck University, the Austrian Science Fund (FWF), and the Austrian Federal Ministry for Science, Research and Economy, the University of Adelaide and the Australian Research Council, the Japan Society for the Promotion of Science and by the University of Amsterdam.
We appreciate the excellent work of the technical support staff in Berlin, Durham, Hamburg, Heidelberg, Palaiseau, Paris, Saclay, and in Namibia in the construction and operation of the equipment. This work benefited from services provided by the H.E.S.S. Virtual Organisation, supported by the national resource providers of the EGI Federation.}

%
%#######################################################################################################################################################################
%_______________________________________________________________________________________________________________________________________________________________________
%

%

\end{document}